\documentclass[12pt,a4wide,epsf]{article}
\usepackage{epsf}

\usepackage{graphicx}

\newcommand{\insertplot}[5]{\begin{figure}
 \hfill\hbox to 0.05in{\vbox to #5in{\vfill
 \inputplot{#1}{#4}{#5}}\hfill}
 \hfill\vspace{-.1in}
 \caption{#2}\label{#3}
 \end{figure}}
 \newcommand{\inputplot}[3]{
 \special{ps: plotfile #1}
}
\newcounter{fig}   

\newcommand{\vphi}{\varphi}
\newcommand{\vepsilon}{\varepsilon}
\newcommand{\DS}{\displaystyle}

\usepackage{graphicx}

\begin{document}

\title{
Rotating 
Einstein-Maxwell-Dilaton\\
Black Holes in D Dimensions
 \vspace{1.5truecm}
\author{
{\bf \hspace{-1cm} Jutta Kunz$^1$},
{\bf Dieter Maison$^2$},
{\bf Francisco Navarro-L\'erida$^1$},
{\bf Jan Viebahn$^1$}\\
$^1$Institut f\"ur Physik, Universit\"at Oldenburg, Postfach 2503\\
D-26111 Oldenburg, Germany\\
$^2$Max-Planck-Institut f\"ur Physik,
F\"ohringer Ring 6\\
D--80805 Munich, Germany\\
}
}

\vspace{1.5truecm}

\date{\today}

\maketitle
\vspace{1.0truecm}

\begin{abstract}
We construct exact charged rotating black holes 
in Einstein-Maxwell-dilaton theory 
in $D$ spacetime dimensions, $D \ge 5$,
by embedding the $D$ dimensional Myers-Perry solutions
in $D+1$ dimensions, and performing a boost 
with a subsequent Kaluza-Klein reduction.
Like the Myers-Perry solutions,
these black holes generically possess $N=[(D-1)/2]$ independent angular momenta.
We present the global and horizon properties of these black holes,
and discuss their domains of existence.
\end{abstract}

\vfill\eject


\section{Introduction}

In $4$ dimensions the Kerr solution presents
the unique family of stationary asymptotically flat vacuum
black holes.
Generalization of these rotating black holes to $D>4$ dimensions 
leads to the Myers-Perry black holes \cite{MP}.
These are characterized by their mass and their $N=[(D-1)/2]$
independent angular momenta, 
associated with $N$ orthogonal planes of rotation 
(where $N=[(D-1)/2]$ denotes the integer part of $(D-1)/2$).
The presence of rotating black rings in $5$ dimensions \cite{blackrings}
however shows, that the uniqueness theorems do not
generalize to higher dimensions.

Coupling the electromagnetic field to gravity yields
the unique family of Kerr-Newman black holes in $4$ dimensions.
Charged black rings \cite{blackrings2} have been
constructed in Einstein-Maxwell (EM) theory in $5$ dimensions,
but no exact solutions of rotating EM black holes with horizons 
of spherical topology are known in $D>4$ dimensions.
So far such black hole solutions have been obtained only numerically 
\cite{KNP}.

In contrast to pure EM theory, exact solutions of higher dimensional
charged rotating black holes are known in theories with more symmetries.
The presence of a Chern-Simons (CS) term, for instance, 
leads to a class of odd-dimensional
Einstein-Maxwell-Chern-Simons (EMCS) theories,
comprising the bosonic sector of minimal $D=5$ supergravity,
whose stationary black hole solutions \cite{BLMPSV,Cvetic}
possess surprising properties \cite{GMT,KN}.
In particular, in EMCS theory,
even black holes with horizons of spherical topology
are no longer uniquely characterized by their global charges
(beyond a critical value of the CS coupling constant)
\cite{KN}.

The inclusion of additional fields, as required by supersymmetry
or string theory, leads to further exact solutions of higher
dimensional black holes \cite{Youm,Horo2},
since then certain constructive methods are available, such as
Hassan-Sen transformations \cite{hassansen}.

Here we construct exact solutions of rotating black holes in $D$ dimensions,
by embedding the $D$-dimensional Myers-Perry solutions
in $D+1$ dimensions, and performing a boost with respect to
the time and the additional coordinate, followed by a
Kaluza-Klein reduction to $D$ dimensions \cite{Maison,emd,rasheed}.
This procedure leads to Einstein-Maxwell-dilaton (EMD) black holes
in $D$ dimensions, for particular values of the dilaton
coupling constant.
The solutions are completely general, with dependence
on the mass $M$, the $N$ angular momenta $\bf J$, and the charge $Q$.
Like their EM counterparts \cite{GMT} they satisfy a Smarr formula
and the first law of black hole mechanics.
The corresponding EMD solutions in 4 dimensions 
have been known since long \cite{emd},
including their dyonic generalizations \cite{rasheed}.

After recalling the Myers-Perry black holes in section 2,
we present the rotating EMD black holes in section 3,
we obtain their global and horizon properties in section 4,
and we discuss their domains of existence in section 5.

\section{Myers-Perry black holes}

The Myers-Perry (MP) solutions of $D$-dimensional gravity
can be presented in a unified way for even and odd dimensional
spacetimes
\cite{frolov}.
For that purpose let us introduce the notation
\begin{equation}
N \equiv \left[\frac{D-1}{2}\right] \ , \vepsilon \equiv \frac{1}{2}(1+(-1)^D)
\ , \label{def_N_eps}
\end{equation}
where $D$ is the dimension of the spacetime 
and $[\cdot]$ denotes the integer part. 
So for even $D$ we have $N=(D-2)/2$ and $\vepsilon=1$,
whereas for odd $D$, $N=(D-1)/2$ and $\vepsilon=0$. 
In what follows we assume $D\geq4$ and choose units such that $16 \pi G_D=1$.

The general MP solutions then read
\begin{eqnarray}
ds_{D,{\rm MP}}^2 &=& -dt^2 + \frac{\Pi F}{\Pi -m r^{2-\vepsilon}} dr^2 + \sum_{i=1}^N
(r^2+a_i^2)(d \mu_i^2 + \mu_i^2 d\vphi_i^2) + \nonumber\\
&& \frac{m r^{2-\vepsilon}}{\Pi F} \left(dt -\sum_{i=1}^N a_i \mu_i^2 d\vphi_i\right)^2+
\vepsilon r^2 d\nu^2 \ , \label{MP_metric}
\end{eqnarray}
where
\begin{equation}
F  \equiv 1 - \sum_{i=1}^N \frac{a_i^2 \mu_i^2}{r^2+a_i^2} \ , \ \ \
\Pi=\prod_{i=1}^N (r^2 +a_i^2) \ . \label{def_F_Pi}
\end{equation}
Note, that the coordinate $\nu$ enters only in even dimensions.
The $\mu_i$ (and $\nu$, for even-dimensional cases) coordinates are not independent but have to obey the constraint
\begin{equation}
\sum_{i=1}^N\mu_i^2 + \vepsilon \nu^2 = 1 \ .\label{constraint}
\end{equation}

\section{Kaluza-Klein black holes}

Let us now generate Kaluza-Klein black holes
using the MP solutions as seeds. 
In order to do so, we first embed the $D$-dimensional MP metric, 
Eq.~(\ref{MP_metric}),
into a $(D+1)$ spacetime with extra coordinate $U$,
\begin{equation}
ds_{D+1}^2 = dU^2 + ds_{D,{\rm MP}}^2 \ , \label{embedded_metric}
\end{equation}
and then perform a boost in the $t-U$ plane with the $2 \times 2$ matrix
\begin{equation}
L=\left(
\begin{array}{cc}
\cosh\alpha&\sinh\alpha \\
\sinh\alpha&\cosh\alpha
\end{array}
\right) \ . \label{boost}
\end{equation}
The resulting boosted $(D+1)$-dimensional metric is a 
solution of the $(D+1)$-dimensional vacuum Einstein equations.

To obtain $D$-dimensional EMD black holes,
the boosted metric has to be compared to the Kaluza-Klein
parametrization of a $(D+1)$-dimensional metric, namely,
\begin{equation}
ds_{D+1}^2=e^{2\iota \Phi} g_{\rho \sigma} dx^\rho dx^\sigma + e^{-2 (D-2)
  \iota \Phi} (dU + A_\rho dx^\rho)^2 \ , \label{KK_paramet}
\end{equation}
where
\begin{equation}
\iota = \frac{1}{\sqrt{2(D-1)(D-2)}} \ , \label{iota}
\end{equation}
and $g_{\rho\sigma}$, $A_\rho$,
and $\Phi$, obtained from this comparison, 
are then identified with the $D$-dimensional
metric, the $D$-dimensional Maxwell potential, 
and the dilaton function, respectively, 
which satisfy the field equations of the $D$-dimensional 
Einstein-Maxwell-dilaton theory with action
\begin{equation}
S=\int d^D x \sqrt{-g} \left(R -\frac{1}{2}\Phi_{,\rho} \Phi^{,\rho} -
  \frac{1}{4} e^{-2 h \Phi} F_{\rho\sigma} F^{\rho\sigma} \right)  \ , \label{action}
\end{equation}
for the particular Kaluza-Klein value of the dilaton coupling constant $h$,
\begin{equation}
h=\frac{D-1}{\sqrt{2(D-1)(D-2)}} \ . \label{h_def}
\end{equation}
These field equations consist of the Einstein equations
\begin{equation}
G_{\rho \sigma} = \frac{1}{2} T_{\rho\sigma} \ , \label{Einstein_eq}
\end{equation}
with stress-energy tensor
\begin{equation}
T_{\rho \sigma}=\partial_\rho \Phi \partial_\sigma \Phi
-\frac{1}{2}g_{\rho\sigma} \partial\tau\Phi \partial^\tau \Phi + e^{-2
  h\Phi}\left(F_{\rho \tau}{F_\sigma}^\tau - \frac{1}{4}g_{\rho \sigma} F_{\tau \beta}F^{\tau
    \beta} \right) \ , \label{se_tensor}
\end{equation}
the Maxwell equations
\begin{equation}
\nabla_\rho \left( e^{-2 h \Phi} F^{\rho \sigma} \right) = 0 \ , \label{Maxwell_eq}
\end{equation}
and the dilaton equation
\begin{equation}
\nabla^2 \Phi = -\frac{h}{2} e^{-2 h \Phi} F_{\rho\sigma} F^{\rho\sigma} \ . \label{dilaton_eq}
\end{equation}

The solution to Eqs.~(\ref{Einstein_eq})-(\ref{dilaton_eq}) then has the
$D$-dimensional metric
\begin{eqnarray}
&&ds_D^2 = g_{\rho\sigma} dx^\rho dx^\sigma = \nonumber \\
&&\left. \left(1+\frac{m
    r^{2-\vepsilon}}{\Pi F} \sinh^2\alpha\right)^\frac{1}{D-2} \right\{  -dt^2
    + \frac{\Pi F}{\Pi-m r^{2-\vepsilon}} dr^2 +  \nonumber \\
&& \sum_{i=1}^N
(r^2+a_i^2)(d \mu_i^2 + \mu_i^2 d\vphi_i^2)+ \vepsilon r^2
d\nu^2 + \nonumber \\
&&\left. \frac{m r^{2-\vepsilon}}{\Pi F + m r^{2-\vepsilon} \sinh^2\alpha}
  \left( \cosh\alpha dt - \sum_{i=1}^N a_i \mu_i^2 d\vphi_i\right)^2 \right\}
\ , \label{metric_solution}
\end{eqnarray}
where $F$ and $\Pi$ are given in Eq.~(\ref{def_F_Pi}) and the constraint
Eq.~(\ref{constraint}) holds.
The gauge potential is given by
\begin{equation}
A_\rho dx^\rho= \frac{m r^{2-\vepsilon}\sinh\alpha}{\Pi F + m r^{2-\vepsilon}
  \sinh^2\alpha}  \left( \cosh\alpha dt - \sum_{i=1}^N a_i \mu_i^2
  d\vphi_i\right) \ , \label{potential_solution}
\end{equation}
and the dilaton function reads
\begin{equation}
\Phi = -\frac{1}{2(D-2)\iota} \log \left(1+\frac{m r^{2-\vepsilon}}{\Pi
    F}\sinh^2\alpha\right)  \ , \label{dilaton_solution}
\end{equation}
with $\iota$ given in Eq.~(\ref{iota}).


\section{Black Hole Properties}

The mass $M$, the angular momenta $J_i$, the electric charge  $Q$, the
magnetic moments ${\cal M}_i$, and the dilaton
charge $\Sigma$ can be read off the
asymptotic behavior of the metric, the gauge potential, and the dilaton
function
\begin{equation}
  g_{tt} = -1 + \frac{M}{(D-3) A(S^{D-2})} \frac{1}{r^{D-3}} +\dots \ , \label{asymp_gtt}
\end{equation}
\begin{equation}
g_{t\vphi_i} = - \frac{J_i}{2A(S^{D-2})} \mu_i^2 \frac{1}{r^{D-3}}  +
\dots \ , \label{asymp_gtphi} 
\end{equation}
\begin{equation}
A_t = \frac{Q}{(D-3)A(S^{D-2})} \frac{1}{r^{D-3}} + \dots \ , \label{asymp_A_t}
\end{equation}
\begin{equation}
A_{\vphi_i} = -\frac{{\cal M}_i}{(D-3)A(S^{D-2})} \mu_i^2
\frac{1}{r^{D-3}} + \dots \ ,  \label{asymp_Aphi} 
\end{equation}
\begin{equation}
\Phi = \frac{\Sigma}{(D-3)  A(S^{D-2})} \frac{1}{r^{D-3}} + \dots \ , \label{asymp_Phi}
\end{equation}
where $ A(S^{D-2})$ is the area of the unit $(D-2)$-sphere.

Comparing these expansions to the asymptotic behavior of the solution,
Eqs.~(\ref{metric_solution})-(\ref{dilaton_solution}), we obtain
\begin{equation}
M=m\left(1+(D-3)\cosh^2\alpha\right) A(S^{D-2}) \ , \label{mass}
\end{equation}
\begin{equation}
J_i = 2 m\, a_i \cosh\alpha A(S^{D-2}) \ , \ \ \ i=1,\dots , N \ ,
\label{angular_momenta}
\end{equation}
\begin{equation}
Q=(D-3)m\,\sinh\alpha\cosh\alpha A(S^{D-2}) \ , \label{electric_charge}
\end{equation}
\begin{equation}
{\cal M}_i = (D-3) m\, a_i \sinh\alpha A(S^{D-2}) \ , \ \ \ i=1,\dots , N \ , \label{magntic_moments}
\end{equation}
\begin{equation}
\Sigma = - \frac{(D-3) m\, \sinh^2\alpha}{2 (D-2) \iota}  A(S^{D-2}) \ . \label{dilaton_charge}
\end{equation}

By combining these global charges one can derive the following quadratic
relation \cite{rasheed}
\begin{equation}
\DS \frac{Q^2}{M -\frac{2(D-2)\iota}{D-3}\Sigma} = - 2 (D-3) \iota \Sigma \
, \label{quadratic_relation}
\end{equation}
which determines the dilaton charge in terms of the mass and the electric
charge. Note that the angular momenta do not enter this relation.

The gyromagnetic ratios $g_i$ are given by
\begin{equation}
g_i = \frac{2 {\cal M}_i M}{Q J_i} = (D-3) + \frac{1}{\cosh^2\alpha}
=g \  , \ \ \ i=1,\dots , N \ . \label{gyromagnetic_ratios}
\end{equation}
The gyromagnetic ratio $g$ depends only
on the charge-to-mass ratio, $q=Q/M$, and ranges between 
$g=D-2$ for $q=0$ \cite{KNP, aliev} and
$g=D-3$ for $|q|=1$ \cite{emd}.

Let us now turn to the horizon properties of these black hole solutions.
Their event horizon is characterized as the
largest non-negative root $r=r_{\rm H}$ of the function
\begin{equation}
\Delta \equiv \Pi -m r^{2-\vepsilon} \ , \label{Delta}
\end{equation}
i.e.,
\begin{equation}
\Delta|_{\rm H} = \left( \Pi -m r^{2-\vepsilon} \right)|_{r=r_{\rm H}}=0 \ , \label{hor_rel}
\end{equation}
where $\Pi$ is given by Eq.~(\ref{def_F_Pi}).
In the general case, Eq.~(\ref{hor_rel}) cannot be solved for $r_{\rm H}$, but
it can be easily solved for $m$, which allows to express certain 
horizon quantities in terms of the horizon radius $r_{\rm H}$.

The (constant) horizon angular velocities $\Omega_i$ can be defined by imposing
the Killing vector field
\begin{equation}
\chi =\partial_t + \sum_{i=1}^N \Omega_i \partial_{\vphi_i} \  \label{chi}
\end{equation}
 to be null on and orthogonal to the horizon, yielding
\begin{equation}
\Omega_i = \frac{a_i}{(r_{\rm H}^2+a_i^2) \cosh\alpha} \ ,  \ \ \ i=1,\dots , N \ . \label{Omegas}
\end{equation}

The area of the horizon $A_{\rm H}$ of these black holes is given by \cite{aman}
\begin{equation}
A_{\rm H} = \frac{\cosh\alpha}{r_{\rm H}^{1-\vepsilon}} A(S^{D-2}) \prod_{i=1}^N
(r_{\rm H}^2 +a_i^2) \ , \label{area_hor}
\end{equation}
and the surface gravity $\kappa_{\rm sg}$, defined by 
\begin{equation}
\kappa_{\rm sg}^2 =
\left. -\frac{1}{2} (\nabla_\mu \chi_\nu) (\nabla^\mu \chi^\nu) 
\right|_{r=r_{\rm H}}
\ , \label{def_surf_gravity}
\end{equation}
takes the form
\begin{equation}
\left. \kappa_{\rm sg} = \frac{\Delta_{,r}}{2 m r_{\rm H}^{2-\vepsilon} \cosh\alpha}
\right|_{r=r_{\rm H}} \ , \label{surf_gravity}
\end{equation}
with $\Delta$ given by Eq.~(\ref{Delta}).

Introducing further the horizon electrostatic potential $\Psi_{\rm el,H}$,
\begin{equation}
\Psi_{\rm el,H} \equiv \chi^\rho A_\rho |_{r=r_{\rm H}}
= \frac{\sinh\alpha}{\cosh\alpha} \ , \label{electros_pot}
\end{equation}
and taking into account the quantities previously defined, it is
straightforward to see that these black holes satisfy the Smarr mass formula
\begin{equation}
M= 2\frac{D-2}{D-3}\kappa_{\rm sg} A_{\rm H} + \frac{D-2}{D-3}\sum_{i=1}^N
\Omega_i J_i + \Psi_{\rm el,H} Q \ . \label{mass_form1}
\end{equation}

Since Eqs.~(\ref{h_def}), (\ref{electric_charge}),
(\ref{dilaton_charge}), and (\ref{electros_pot}) yield the
relation
\begin{equation}
\frac{\Sigma}{h} = - \Psi_{\rm el,H} Q \ , \label{dilaton_rel}
\end{equation}
Eq.~(\ref{mass_form1}) leads to the modified Smarr-like formula
\begin{equation}
M= 2\frac{D-2}{D-3}\kappa_{\rm sg} A_{\rm H} + \frac{D-2}{D-3}\sum_{i=1}^N
\Omega_i J_i + 2\Psi_{\rm el,H} Q + \frac{\Sigma}{h}\ , \label{mass_form2}
\end{equation}
also valid for non-Abelian black holes (in $D=4$) \cite{kkrot}.

\section{Domain of existence}

In $D=4$ dimensions the Kerr black holes satisfy the relation
\hbox{$M^2\ge \mid 16\pi J_1\mid$}, while for the Kerr-Newman black holes
of EM theory $M^2\ge 4 Q^2+(16 \pi J_1)^2/M^2$ holds, ensuring cosmic censorship.
The bounds are saturated for extremal solutions,
which thus enclose the domain of existence of EM black hole solutions,
exhibited in Fig.~1.
\begin{figure}[h!]
\parbox{\textwidth}
{\centerline{
\mbox{
\epsfysize=10.0cm
\includegraphics[width=70mm,angle=0,keepaspectratio]{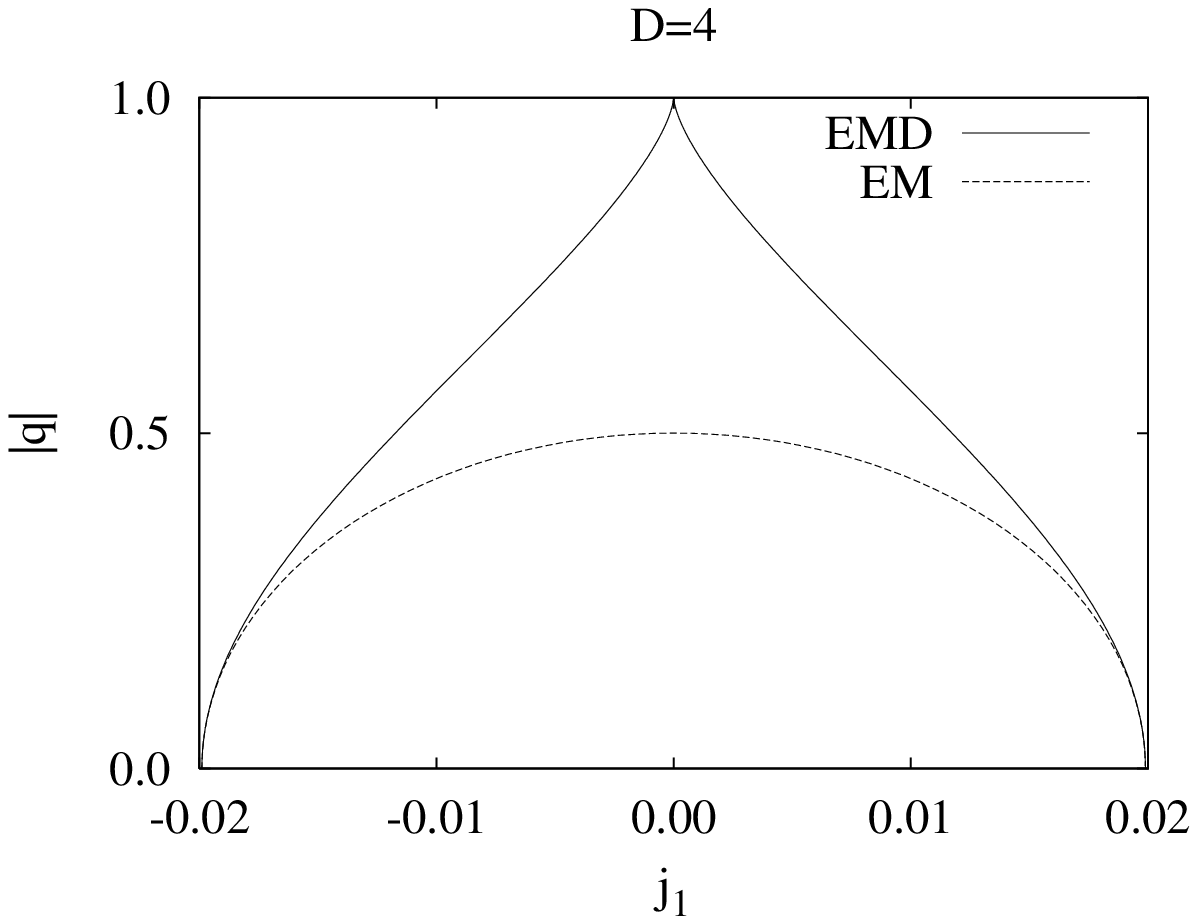}
\includegraphics[width=70mm,angle=0,keepaspectratio]{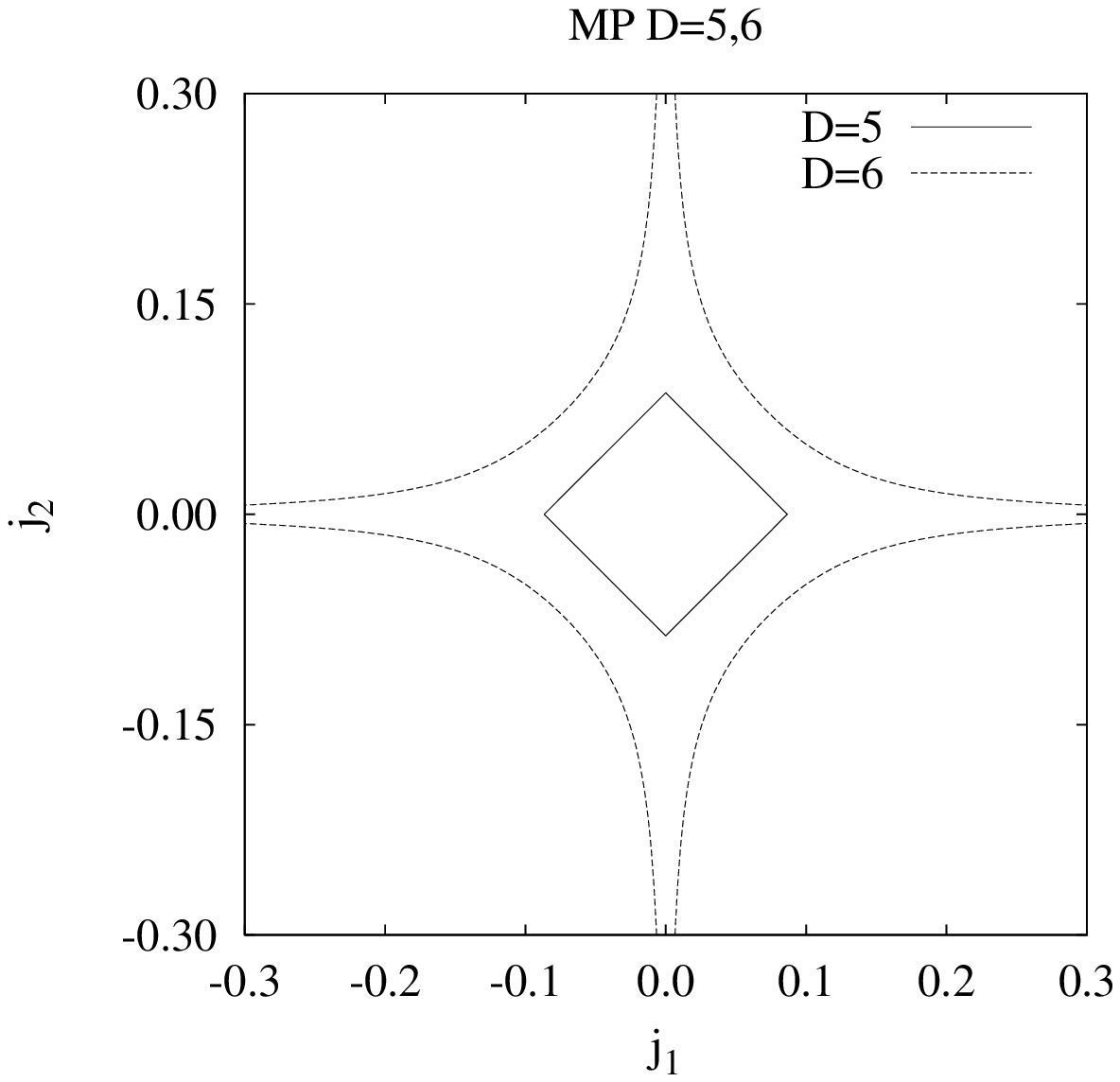}
}}}
\caption{
Left:
Domain of existence of charged black holes in $D=4$:
scaled charge $q=Q/M$ versus scaled
angular momentum $j_1=J_1/M^{2}$ 
for extremal EM and EMD solutions.
Right:
Domain of existence of MP black holes in $D=5$ and $D=6$:
scaled angular momentum $j_1=J_1/M^{(D-2)/(D-3)}$ versus
$j_2=J_2/M^{(D-2)/(D-3)}$
for extremal solutions.
}
\end{figure}

The 4-dimensional EMD black holes satisfy the relation 
$M^2 \ge (2 Q)^2 + (16 \pi a_1)^2 - (2 \Sigma)^2$
\cite{emd,rasheed},
which is precisely the condition for the original Kerr solution to have
horizons. 
The domain of existence of these EMD black holes
is also exhibited in Fig.~1 and contains the EM domain.
Unlike the EM case,
the extremal EMD solutions do not form a smooth boundary, 
but form a cusp at $J=0$, 
and the associated static extremal solutions
possess vanishing horizon area. In contrast, rotating extremal
solutions possess finite horizon area \cite{emd,foot2}.

Considering the domain of existence of $D=5$ MP black holes,
a similar feature is observed. 
$D=5$ black holes possess two independent angular momenta.
The extremal $D=5$ MP solutions then
form a square with respect to the scaled angular momenta
$j_i=J_i/M^{3/2}$, $i=1,2$,
as seen in Fig.~1. At the vertices of the square
one of the two angular momenta vanishes, 
and the associated extremal single angular momentum solutions
possess vanishing horizon area. Generic extremal
solutions (with two non-vanishing angular momenta),
in contrast, possess finite horizon area.

The domain of existence of the $D=5$ EMD black holes is exhibited
in Fig.~2.
We observe, that extremal $D=5$ EMD solutions
retain the main features of extremal $D=5$ MP solutions.
Thus for a given scaled charge $q=Q/M$, the extremal EMD solutions
form a square, and have vanishing horizon area at the vertices.
With increasing $|q|$ the square shrinks and reaches zero size for
$|q|=1$.
\begin{figure}[h!]
\parbox{\textwidth}
{\centerline{
\mbox{
\epsfysize=10.0cm
\includegraphics[width=90mm,angle=0,keepaspectratio]{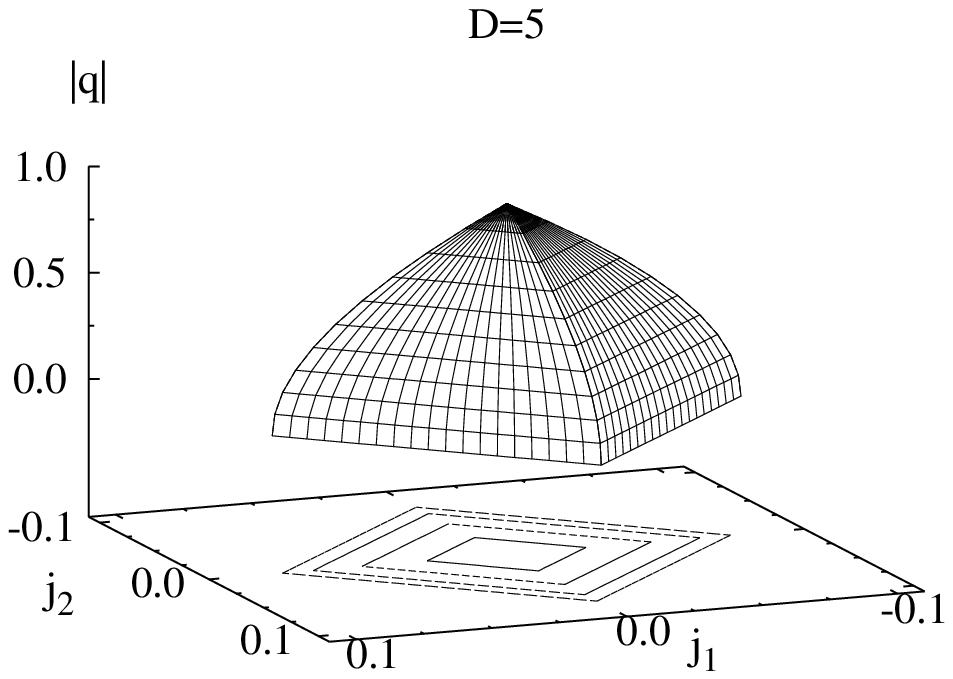}
\includegraphics[width=90mm,angle=0,keepaspectratio]{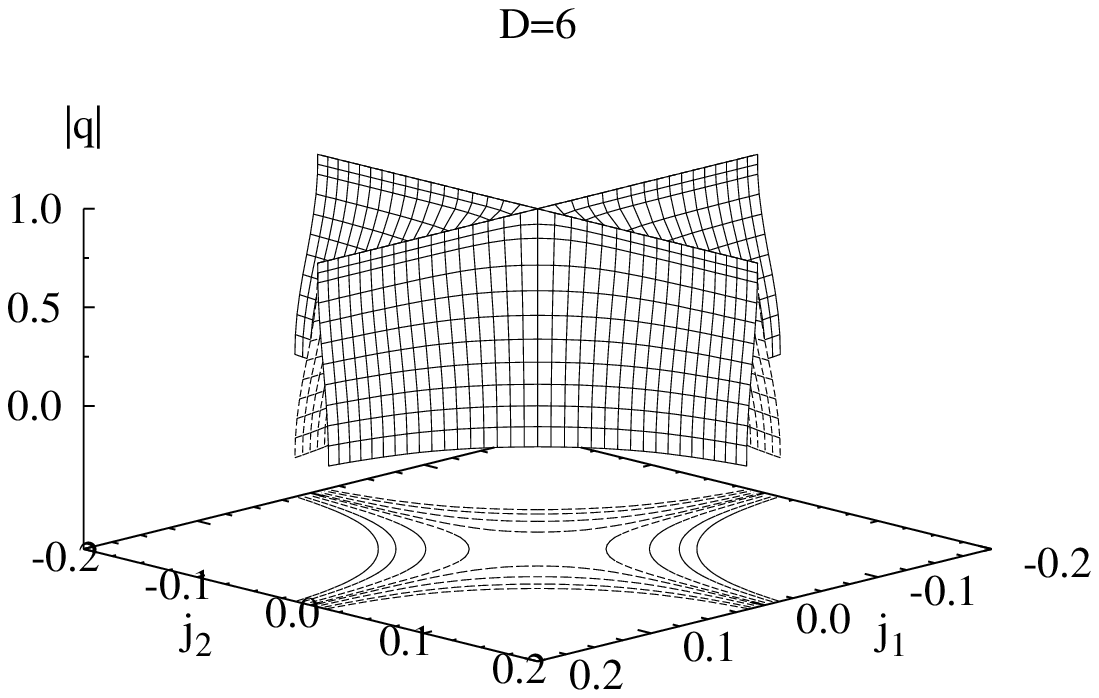}
}}}
\caption{
Domain of existence of EMD black holes in $D=5$ (left) and $D=6$ (right):
scaled charge $|q|=|Q|/M$ 
versus scaled angular momenta $j_i=J_i/M^{(D-2)/(D-3)}$, $i=1,2$
for extremal EMD solutions.
}
\end{figure}

When moving to $D=6$ dimensions,
the domain of existence of MP solutions changes distinctly.
The vertices present in the $D=5$ domain then move to infinity,
while the boundary lines between the vertices are no longer straight
but are given by 
\begin{equation}
\DS 1=\frac{256 \sqrt{6}}{27} \pi^2  \left[\sqrt{(j_1^2+j_2^2)^2+12
    j_1^2 j_2^2} + 2 (j_1^2+j_2^2) \right] \sqrt{ \sqrt{(j_1^2+j_2^2)^2+12
    j_1^2 j_2^2}-(j_1^2+j_2^2)} \ .
\label{6D_extremal_limit}
\end{equation}
Thus there are no extremal solutions
for single angular momentum black holes. 
In contrast, generic black holes with two non-vanishing
angular momenta possess extremal limits,
as illustrated in Fig.~1.

The domain of existence of the $D=6$ EMD black holes is exhibited
in Fig.~2.
Again, the extremal $D=6$ EMD solutions
retain the main features of extremal $D=6$ MP solutions.
Thus for a given scaled charge $q=Q/M$, the extremal EMD solutions
form hyperbola-like curves, diverging on the $j_1$ and $j_2$ axes.
With increasing $|q|$ the curves approach the $j_1$ and $j_2$ axes,
reaching them in the limit $|q|=1$.
The solutions on these axes have vanishing area.

This $D=6$ pattern is now retained and generalized
when going to higher dimensions.
In $D>6$ dimensions generic black holes with $N=[D-1]/2$
non-vanishing angular momenta
are delimited by extremal solutions.
The domain of existence is unbounded though, since the $j$-axes
are part of it. 
The inequality $|q| \le 1$ remains valid for any dimension $D>6$. 
As $|q|$ increases, the $N$-dimensional volume of black hole solutions
for constant $q$ shrinks in size. In the limit $|q|=1$,
the resulting surface of extremal EMD solutions contains
the $j$-axes, as illustrated for $D=7$ in Fig.~3.
\begin{figure}[h!]
\parbox{\textwidth}
{\centerline{
\mbox{
\epsfysize=10.0cm
\includegraphics[width=90mm,angle=0,keepaspectratio]{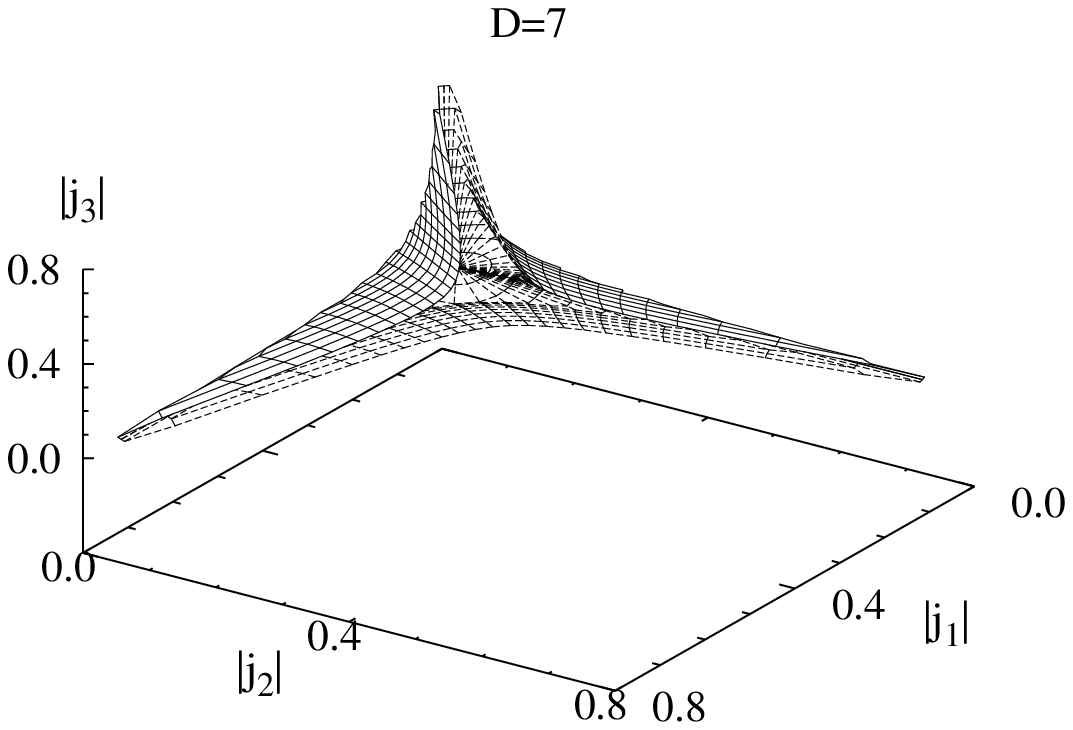}
\includegraphics[width=90mm,angle=0,keepaspectratio]{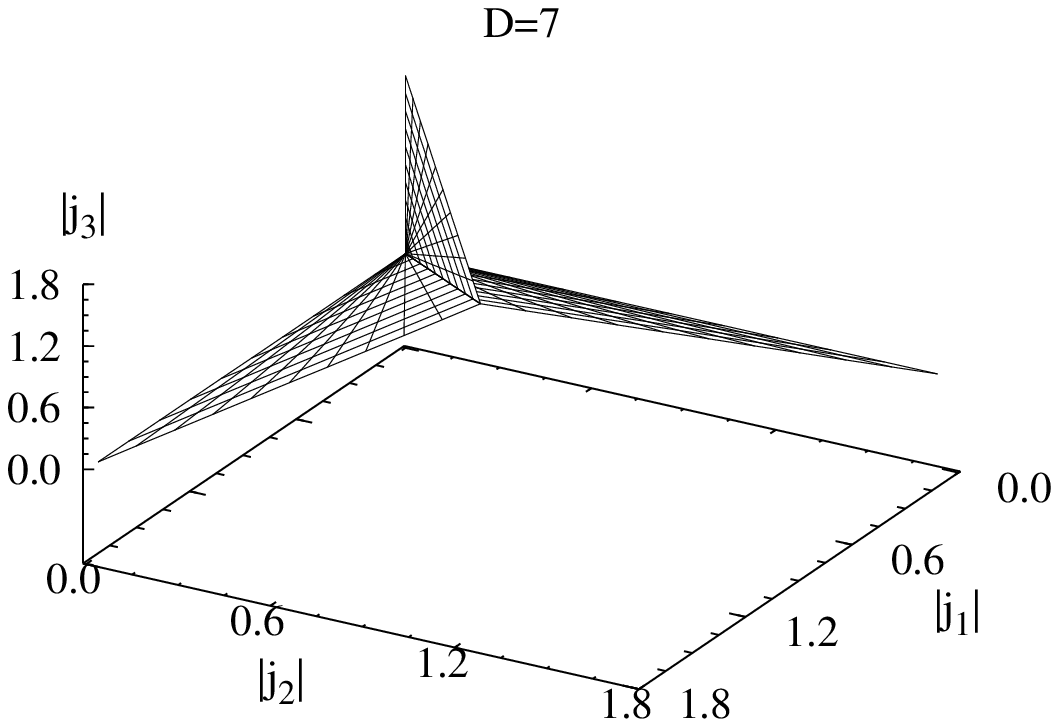}
}}}
\caption{
Left: Domain of existence of MP black holes in $D=7$
in terms of scaled angular momenta $j_i=J_i/M^{(D-2)/(D-3)}$, $i=1,2,3$.
Right: Extremal EMD black holes in $D=7$
for the extremal value of the scaled charge $|q|=1$.
}
\end{figure}

\section{Conclusions}

Based on the Kaluza-Klein action in $D$ dimensions,
obtained by reducing the $D+1$ dimensional vacuum Einstein action,
we have constructed exact rotating black hole solutions
by embedding the $D$ dimensional MP solutions in $D+1$ dimensions
and boosting in the extra direction.

The resulting black hole solutions are asymptotically flat, 
and possess a regular horizon of spherical topology.
They are characterized by their global charges: their mass,
their $N=[(D-1)/2]$ angular momenta, and their electric charge.
Their dilaton charge is not independent, but determined by their
mass and electric charge, Eq.~(\ref{quadratic_relation}).
Their gyromagnetic ratio covers the range $D-3 \le g \le D-2$.

Combining their global charges and their horizon properties,
these rotating black holes are seen to satisfy the first law
and a Smarr formula Eq.~(\ref{mass_form1}), where the electric
charge term can be replaced by a dilaton charge term,
Eq.~(\ref{mass_form2}).

The domain of existence of these black holes is obtained
by considering the set of extremal black holes.
In 5 dimensions there exist always extremal solutions,
but when one of the two angular momenta vanishes,
these extremal solutions possess vanishing horizon area.
In higher dimensions, for $|q|<1$  extremal solutions do not exist,
when one (for even $D$) or two (for odd $D$) of the angular momenta vanish.
In the limit $|q|=1$ the $j$-axes are contained in the surface of
extremal EMD solutions.

Here these higher dimensional rotating EMD black hole solutions 
were obtained only for particular values of the dilaton
coupling constant. It remains a challenge to generalize these solutions
to arbitrary values of the dilaton coupling constant,
including the pure Einstein-Maxwell case \cite{emdring}.

{\bf Acknowledgement}

FNL gratefully acknowledges Ministerio de Educaci\'on y Ciencia for
support under grant EX2005-0078.


\begin{thebibliography}{000}
\bibitem{foot2}
 For extremal static EMD solutions with both electric and magnetic charge
 the horizon area is finite \cite{rasheed}.
\end{thebibliography}
\end{document}